\documentclass[elsart]{elsart}
\usepackage{graphicx,amssymb,amsmath,times}
\journal{Astroparticle Physics}
\begin{document}

\begin{frontmatter}

\title{Search for diffuse cosmic gamma-ray flux using Fractal and Wavelet analysis from Galactic region using single imaging Cerenkov telescopes}
\author{C.K.Bhat}
\address{Astrophysical Sciences Division, Bhabha Atomic Research Centre\\ Mumbai-400085, India}
\ead{ckbhat@barc.gov.in} 
\begin{abstract}
{We show from a simulations-based study of the TACTIC telescope that fractal and wavelet analysis of Cerenkov images, recorded in a single imaging Cerenkov telescope, enables almost complete segregation of isotropic gamma-ray initiated events from the overwhelming background of cosmic-ray hadron-initiated events. This presents a new method for measuring galactic and extragalactic gamma-ray background above 1 TeV energy. Preliminary results based on this method are reported here.Primary aim is to explore the possibility of using data recorded by a single imaging  atmospheric Cerenkov telescope$(IACT)$ for making accurate measurements of diffuse galactic and extragalactic gamma-ray flux above $\sim$ 1 TeV energy.Using simulated data of atmospheric Cerenkov images recorded in an IACT, initiated , both, by cosmic ray protons and diffuse gamma-rays with energies above 4 TeV and 2 TeV respectively, we identify the most efficient fractal$\diagup$wavelet parameters of the recorded images for primary identification. The method is based on the pattern recognition technique and employs fractal and wavelet analysis of the recorded Cerenkov images for gamma-hadron segregation.We show that the value of wavelet dimension parameter $B_{6}$ can segregate Cerenkov images of hadronic origin from those of diffuse gamma-ray origin with almost 100\%\ accuracy. We use the results to get a preliminary upper limit estimate of the diffuse galactic gamma-ray flux within galactic range of $|b|$$\leq$ -5$^{0}$ and $|l|$$\leq$ 200$^{0}$ above 2 TeV from a 36h data set recorded by the TACTIC telescope.}
\end{abstract}

\end{frontmatter}

\section{Introduction}
%\label{}
The diffuse high energy gamma-ray background from the galactic plane, mapped extensively at energies above 100 MeV by the EGRET experiment, is generally attributed to nucleon-nucleon scattering and the decay of neutral pions produced in cosmic ray interactions with the interstellar matter (ISM)[1-5]. As the galactic cosmic ray spectrum, from sources like SNR and pulsars,is expected to extend upto $\sim$ 10$^{15}$ eV, the diffuse gamma-ray spectrum resulting from cosmic ray interactions with the ISM can easily extend into the 10$^{'}$s of TeV range. Imaging atmospheric Cerenkov telescopes, which operate at energies of $\sim$ 1 TeV, can look for galactic diffuse gamma-ray emission especially if two or more such systems are operated in the stereoscopic mode to enable accurate primary arrival direction reconstruction. However, imaging Cerenkov telescopes have an inherently low sensitivity for differentiating between background cosmic-ray initiated events and the off-axis(diffuse) gamma-ray initiated events because the image shape and orientation parameters only separate primary gamma ray-initiated events with arrival direction along the principal axis of the light collector[6]. There have only been a few reports of searches for diffuse galactic gamma-ray emission using imaging Cerenkov telescopes and in several cases upper limits on the flux at energies $\geq$ 1 TeV have been placed [7-10]. The first unambiguous detection of $\geq$ 1TeV diffuse gamma-rays from the galactic plane has recently come from the MILAGRO experiment. which is a wide angle water Cerenkov detector with excellent angular resolution [11]. In view of the importance of galactic diffuse gamma-ray measurements at TeV energies and the availability of several imaging Cerenkov systems, we have conducted a detailed simulations- based study of Cerenkov images recorded in the TACTIC telescope to identify additional non-imaging parameters which have the potential to provide clear segregation between events initiated by isotropic gamma-rays and those of cosmic ray hadron origin. Here we report our success in segregating these two types of events, recorded by an imaging telescope, by exploiting the fractal and wavelet characteristics of the image and use our results to obtain a preliminary estimate of the galactic diffuse gamma-ray flux at $\geq$ 1 TeV energy from  a 36h long TACTIC data set.
\section{TACTIC system}
%\label{}
The TACTIC ( TeV Atmospheric Cerenkov Telescope with Imaging Camera ) gamma-ray telescope is operating at Mt. Abu ( 24.6$^{0}$ N,72.7$^{0}$ E, 1300  m asl), India, for the last several years[12] . It essentially comprises a light collector (F/1 type) of $\sim$ 9.5m$^{2}$ area which is made up of 34$\times$ 0.6m diameter,front coated spherical glass mirrors prealigned to produce an on-axis spot size of 0.3$^{0}$ diameter at the focal plane. The telescope is equipped with a 349-pixel photomultiplier tube-based imaging camera with a field of view of 6$^{0}$$\times$6$^{0}$ and uniform pixel resolution of$\sim$ 0.31$^{0}$. The inner most 225(15$\times$15) pixels are used for generating the event trigger based on the 3NCT (Nearest Neighbour, Non-Collinear Triplet) topological logic[13] . Operating at a $\gamma$-ray threshold energy of $\sim$ 1 TeV, the telescope records the Cerenkov images of gamma-ray and cosmic ray hadron-initiated showers with a typical rate of $\sim$ 2Hz in the vertical direction. The imaging data is subjected to standard image cleaning procedure before being parameterized in terms of the standard image shape and orientation parameters.A detailed description of TACTIC telescope and the data reduction procedures is available in [12].
\section {Simulations}
The CORSIKA (version 5.61) air shower simulation code, developed by the KASCADE group[14-16] has been used for simulating the extensive air shower development and production of Cerenkov photons in the atmosphere at Mt. Abu by primary cosmic-ray protons and gamma-rays. A total of 6000 showers  each have been generated for cosmic gamma-ray primaries with energies 2, 5, 10, 20, 30 and 50 TeV and an equal number for cosmic ray protons with energies 4, 10, 20, 40,60 and 100 TeV respectively. The primary particles in both the cases have been considered to be isotropically incident, within the trigger field of view of 4$^{0}$$\times$ 4$^{0}$ of the TACTIC telescope, from a direction making an angle of 50$^{0}$ with the vertical. The high value of zenith angle has been deliberately chosen to take advantage of the larger effective collection area(alongwith a proportionate increase in the threshold energy) available in the large zenith angle mode of operation[17]. Atmospheric extinction of Cerenkov photons has been taken into account using appropriate wavelength-dependent extinction coefficients and the photon distribution in the 349-pixel camera derived after appropriate ray tracing of the photons for all events passing the 3NCT trigger criterion[18] . An additional data set of atmospheric Cerenkov events initiated by primary gamma-rays of 2 TeV energy, incident from a zenith direction of 20$^{0}$along the principal axis of the TACTIC reflector has been compiled for comparison with TACTIC on -source observations.The derived Cerenkov images have been subjected to the usual flat-fielding, image cleaning and calibration procedures before being subjected to the fractal and wavelet analysis as described below.
\section{Comparison with TACTIC data}
%\label{}
In order to check whether the simulations reproduce exactly the expected behaviour of the TACTIC telescope, we have compared the imaging analysis results with results from actual on and off source data scans respectively.The image parameters have been calculated as per the procedure described in[19] . It needs to be kept in mind that a predominant majority of Cerenkov events recorded by the TACTIC telescope in both the on-and off source scans is of cosmic ray origin(mainly protons)while the on-source scan may also contain a possible small gamma-ray signal from the source which is being tracked by the telescope. Fig.1 shows a comparison of the distribution of image parameters L,W,D and Alpha for simulated 4 TeV energy cosmic ray proton initiated events and an equal number of events from a TACTIC off-source scan recorded during a Crab nebula observation compaign.As expected , the image parameter distributions for the two samples completely overlap each other showing that the simulations are able to reproduce the expected behaviour for cosmic ray proton events. This conclusion is reinforced by the results presented in Fig.2 where we show a similar comparison between gamma-ray like events selected by using the prescribed imaging cuts to a TACTIC on -source data set comprising 259619 events and the results from an equal number of simulated 2 TeV gamma-ray events incident along the principal axis of the light collector inclined at an angle of 20$^{0}$with respect to the vertical[20]. Since the majority of the events in the experimental data set again result from cosmic ray protons,erroneously classified as gamma-ray like events from their image parameter values,we expectedly find a much poorer agreement between the two distributions even though the Crab on-source data contains a statistically significant gamma-ray signal as revealed in the Alpha distribution plot. The above tests conducted with the on and off source data sets and the simulated data showing that the simulations are able to reproduce exactly the expected behaviour of the TACTIC telescope,thus validate the use of the simulated data sets for search of other non-imaging parameters for gamma/hadron segregation as described below.   
%-----------------------------------------------------------              % S_vib
   \begin{figure}
   \centering 
    \includegraphics*[width=0.60\textwidth,angle=0,clip]{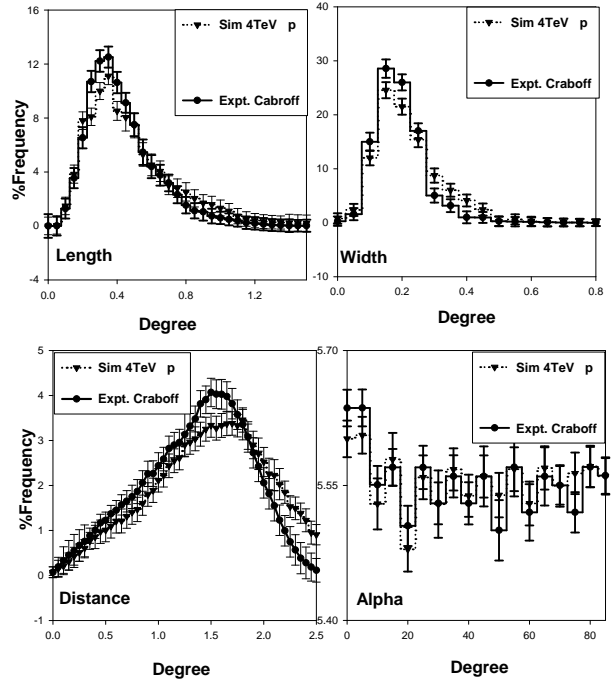}
    \caption{\label{Fig.1}Comparison of image parameter distribution (L,W,D and Alpha) for simulated 4 TeV proton-initiated events and actual TACTIC data recorded during a Crab off-source scan}
    \end{figure} 

% \begin{figure}
%   \centering
%     \includegraphics[width=9cm,height=10cm]{newpafig1.eps}
%  \caption{\label{fig1}Comparison of image parameter distribution (L,W,D and Alpha) for simulated 4 TeV proton-initiated events and actual TACTIC data recorded during a Crab off-source scan.
%         } 
%         \label{FigVibStab}
%   \end{figure}
%______________________________________________________________
\section{Fractal and Wavelet analysis of Cerenkov images}
%\label{}
The pattern recognition technique for gamma-hadron separation employs fractal and wavelet analysis of Cerenkov images to exploit differences in the structure of the recorded images due to differences in the longitudinal and lateral development of showers initiated by the two primary species. Showers initiated by primary gamma-rays and protons produce different numbers of relativistic electrons and muons and also have different heights (and lateral extents) of shower maximum. These differences lead to structures in the Cerenkov image on different scale lengths which can be parameterized through multifractal and wavelet dimensions and moments  of  different order[21-23] . The usual procedure for calculating these parameters involves dividing the Cerenkov image into M = 4, 16, 64, 256 equally sized, non-overlapping cells and finding the number of ADC counts in each cell. The multifractal moments of order q(q = 2$-$ 6) are calculated from the relation 

\begin{eqnarray}
\textrm{Fractal moment} &G_{q}(M) &= \sum_{j=1}^{M}\left[\frac{K_{j}}{N}\right]^{q} 
\end{eqnarray}

where N is the total number of ADC counts in the image, $K_{j}$ is the ADC count in the j$^{th}$ cell in particular scale and $K_{j+1}$ in the j$^{th}$ cell in the consecutive scale and q= 2, 3, 4, 5, 6 is the order of fractal moment. The wavelet moments are obtained from
\begin{eqnarray}
\textrm{Wavelet moment} &W_{q}(M)&=\sum_{j=1}^{M}\left[(\frac{K_{j+1}-K_{j}}{N})\right]^{q} 
\end{eqnarray}

The fractal scale length is given by ${\nu}$ = $log_{2}$M 

Both the fractal moment $G_{q}$ as well as the wavelet moment $W_{q}$ exhibit a proportionality with the scale length which can be expressed as,

\begin{equation}
  G_{q}(M) \propto M ^{\tau_q}\qquad\textrm{and}\qquad\ 
  W_{q}(M) \propto M ^{\beta_q} 
\end{equation}
    
The fractal and wavelet dimensions of order q are then determined from the relations
 
\begin{eqnarray}
\textrm{Fractal dimension}   &D_{q} &= \frac{\tau_q}{q-1} \\
\textrm{Wavelet dimension}   &B_{q}&= \frac{\beta_q}{q-1}
\end{eqnarray}

where $\tau_{q}$ and $\beta_{q}$ are slopes of the line drawn between natural logarithm of $G_{q}$(M) and $W_{q}$(M) with scale length $\nu$ respectively. Since wavelet dimensions are more sensitive to local structures in the photon distribution in Cerenkov images, they are expected to have better efficiency for gamma-hadron separation as proton-initiated showers are known to have a non-uniform structure due to contributions from pions and muons[17].
\section{Results from simulations}
%\label{}
Fig.3 and Fig.4 show a comparison of the frequency distributions of fractal and wavelet dimensions $D_{2}$,$B_{2}$,$D_{6}$ and $B_{6}$ for simulated Cerenkov images produced by isotropic( within TACTIC field- of- view) 2 TeV gamma and 4 TeV proton- initiated atmospheric Cerenkov events. Also shown for comparison are the $B_{2}$,$D_{2}$ and $B_{6}$,$D_{6}$ distributions for a TACTIC off- source data set of thirty six hour duration. For q$\geq$ 6 both the fractal and the wavelet moments exhibit saturation as a function of scale length so that the fractal and wavelet dimensions $D_{2}$, $D_{6}$,$B_{2}$ and $B_{6}$ represent the two extreme values of these parameters which can be tested for gamma-hadron separation.It has been pointed out earlier[22]that the fractal dimensions $D_{2}$ and $D_{6}$ distinguish between images containing a single large peak in the centre and images with several peaks or a smooth distribution while the wavelet dimensions examine differences in neighbouring pixels at different length scales.
%-----------------------------------------------------------                   S_vib
   \begin{figure}
   \centering 
    \includegraphics*[width=0.60\textwidth,angle=0,clip]{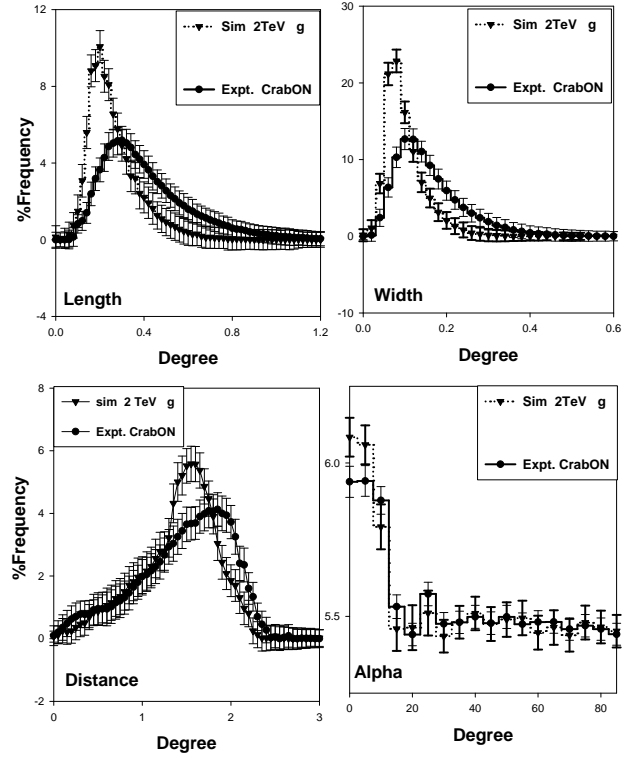}
    \caption{\label{Fig.2}Comparison of image parameter distribution (L,W,D and Alpha)
for simulated 2 TeV gamma-ray initiated events(incident along the reflector axis from 20$^{0}$ zenith direction) with actual gamma-ray like events extracted from a TACTIC on-source scan of Crab nebula.}
    \end{figure}  

%   \begin{figure}
%     \centering
%     \includegraphics[width=9cm,height=10cm]{newpafig2.eps}
%     \caption{\label{fig2}Fractal and wavelet dimension distributions of Cerenkov
%      images generated by gammas and hadrons(p,Ne,Fe)} 
%    \end{figure}

% \begin{figure}
%   \centeringnewpaBlafig2.eps
%     \includegraphics[width=9cm]{newpafig2.eps}
%   \caption{Comparison of image parameter distribution (L,W,D and Alpha)
%for simulated 2 TeV gamma-ray initiated events(incident along the reflector axis from 20$^{0}$ zenith direction) with actual gamma-ray like events extracted from a TACTIC on-source scan of Crab %Nebula.
%      }
%         \label{FigVibStab}
%   \end{figure}

%______________________________________________________________
We find from Fig.3 and Fig.4 that in the case of fractal dimensions the frequency distributions of $D_{2}$ and $D_{6}$ for gamma-ray and proton-initiated events are very broad and overlap over most of the range except for a small peaked structure at $D_{2}$= 0.60 and $D_{6}$= 0.50 for gamma-ray events. However, in the case of wavelet dimensions $B_{2}$ and $B_{6}$, we find that the frequency distributions for gamma and proton-initiated events are well separated from each other. The best segregation is observed for the parameter $B_{6}$  where the frequency distribution for proton-initiated events is found to be strongly peaked at $B_{6}$ = 1.75 while the frequency distribution of $B_{6}$for gamma-initiated events produces a slightly broader peak with a prominent maximum at $B_{6}$ = 5.75. For each of the four parameters we have chosen a domain of parameter values which passes the maximum number of gamma-ray initiated events and rejects the largest number of cosmic ray proton initiated events to quantify the gamma-hadron segregation capability. The efficiency for gamma-hadron segregation has been quantified in terms of the quality factor Q in each case from the relation.
         
              \begin{displaymath}
               \mathrm{Q}=                             
               \frac{N_{\gamma}}{N_{\gamma T}}
               \left(
               \frac{N_p}{N_{pT}}\right)^{-1/2}
               \end{displaymath}

where $N_{\gamma}$ is the number of gamma ray events accepted out of the total number ($N_{\gamma T})$ of gamma-initiated events after applying the parameter cut. $N_{p}$represents the number of CR proton events retained out of the total ($N_{p T}$) proton-initiated events after applying the parameter domain cut. The highest quality factor (Q=8) is obtained in the case of $B_{6}$,suggesting that this parameter can provide $\geq$ 99\%\ gamma-hadron segregation. As can also be seen visually from Fig.4, $\geq$ 90\%\ proton events have $B_{6}$$\leq$ 2, while $\geq$ 90\%\ gamma-ray initiated events have $B_{6}$$\geq$ 5, indicating the high efficiency for gamma-hadron segregation using this parameter.
%--------------------------------------------------------------
   \begin{figure}
   \centering
      \includegraphics*[width=0.60\textwidth,angle=0,clip]{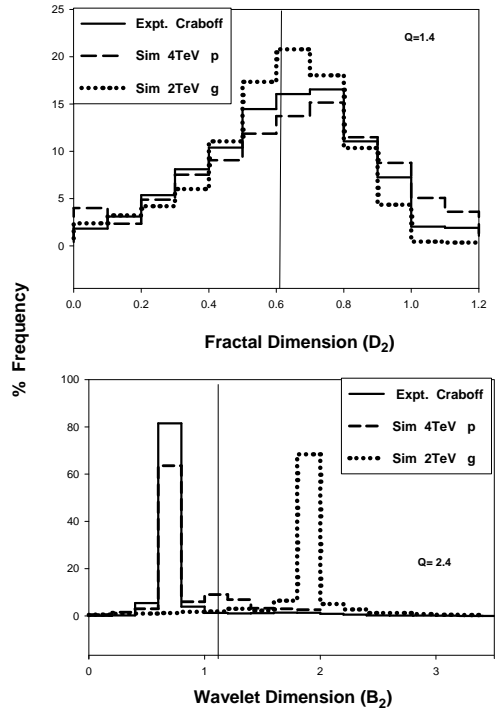}
      \caption{\label{Fig.3}Comparison of distributions for Fractal and Wavelet dimension parameters $D_{2}$ and $B_{2}$ for simulated 4 TeV proton initiated events and 2 TeV isotropic gamma-ray initiated events. The vertical lines represent the parameter cut-off value used to estimate the gamma-proton segregation efficiency($\sim Q$). For details refer to the text.}
%         \label{FigVibStab}
   \end{figure}

%______________________________________________________________

 \begin{figure}
   \centering
     \includegraphics*[width=0.60\textwidth,angle=0,clip]{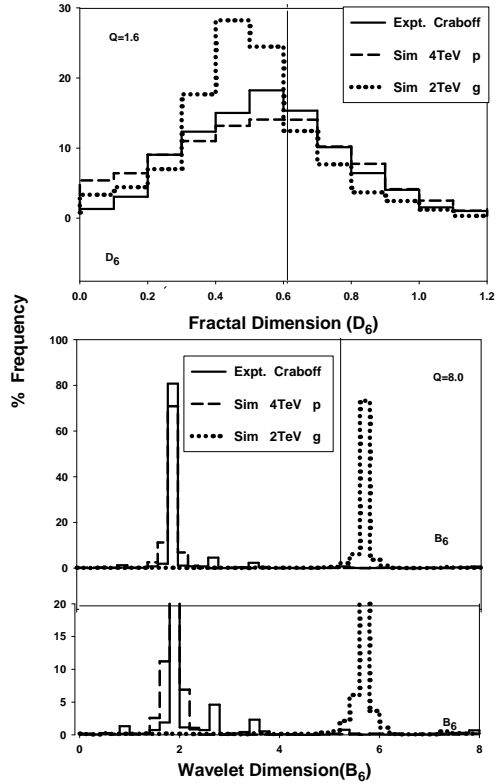}
      \caption{\label{Fig.4}Comparison of distributions for Fractal and Wavelet dimension parameters $D_{6}$ and $B_{6}$ for simulated 4 TeV proton initiated events and 2 TeV isotropic gamma-ray initiated events and actual data recorded during a 36h off-source scan. The lowest panel of figure shows an expanded version of the lower part of $B_{6}$ distribution. The vertical lines represent the parameter cut-off value used to estimate the gamma-proton segregation efficiency($\sim Q$). For details refer to the text.}
%         \label{FigVibStab}
   \end{figure}

%______________________________________________________________
\section{Experimental results} 
%\label{}
The TACTIC telescope has been observing the Crab nebula and a few extragalactic sources regularly since 2000 in the on-source mode. Occasionally in order to obtain a realistic estimate of the background, the telescope is used to track an off-source region which is a region of the sky close ($\sim$ few degrees away in declination but at the same right ascension)to the candidate source and is known to be devoid of any TeV gamma-ray source.We have subjected several Crab off-source data sets of $\sim$ 36h duration in all to the above fractal and wavelet analysis to isolate the background diffuse gamma-ray events(if any) from the recorded events using the $B_{6}$ parameter cut. Fig.5 shows the frequency distribution of $B_{6}$ for the events recorded during such off-source scans from a zenith direction of $\sim$ 50$^{0}$ with 3h $\leq$ $\alpha$ $\leq$ 7h and $\delta= $ 22$^{o}.00min.52sec$ as off-source coordinates. The total number of events recorded during the Crab off-source scans was 259600, corresponding to an event rate of $\sim$2.0 Hz, which translates into a threshhold detection energy of 4 TeV for cosmic ray proton-initiated events ($\sim$2 TeV for gamma-ray initiated events) as already considered in the simulation exercise. While 220320$\pm$ 469  events out of the total events $\sim$ 85\%\ fall within the proton domain of $\leq$ 2 ,only 154 $\pm$ 13 events fall in the gamma-ray domain of $\geq$ 5. Assuming the gamma domain events to result from diffuse galactic gamma-rays and taking into account the exposure time (36h) TACTIC effective area (7.07$\times$ 10$^{8}$ cm$^{2}$) and the TACTIC field of view(3.8$\times$ 10$^{-3}$sr),the upper limit of diffuse gamma ray flux ($\phi^{UL}$) above 2 TeV within 40\%\ systematic error [20]is found to be 
$(2.4\pm 0.3)$10$^{-10}$\textrm{photons\quad}cm$^{-2}s^{-1}sr^{-1}$.This upper limit is plotted in Fig.6.
%\begin{equation}
%J_{\gamma} ( > 2 TeV)=(2.4\pm 0.3)10^{-10}\textrm{photons\quad}cm^{-2}s^{-1} sr^{-1}
%\end{equation}

Although this preliminary result is based on simulation-based fractal parameter cuts which are yet to be optimized, the above estimate for galactic diffuse gamma-ray flux above 2 TeV compares favourably with the upper limit of 3$\times$10$^{-8}$cm$^{-2}$s$^{-1}$sr$^{-1}$ above 0.5 TeV reported by the Whipple group[9]and $\leq$1.96$\times$10$^{-10}$ cm$^{-2}$s$^{-1}$ sr$^{-1}$  above 10 TeV reported by the Tibet air shower group[24]. Fig.6 also shows the measurements and upper limits obtained by other experiments observing the Galactic region in the energy range below and beyond 2 TeV. Analysis of simulated data has revealed that fractal and  wavelet parameters has the capability of better segregation of the cosmic-ray and gamma-ray initiated Cerenkov events. In order to test this hypothesis, we used fractal and wavelet analysis approach on one spell of 66h Crab nebula on-source observation data collected during Dec2007-Jan08 using TACTIC telescope and results obtained here showed better matching of experimental and simulated non-image parameters($B_{6}$,$D_{6}$)[25].         

\begin{figure}
  \centering
     \includegraphics*[width=0.60\textwidth,angle=0,clip]{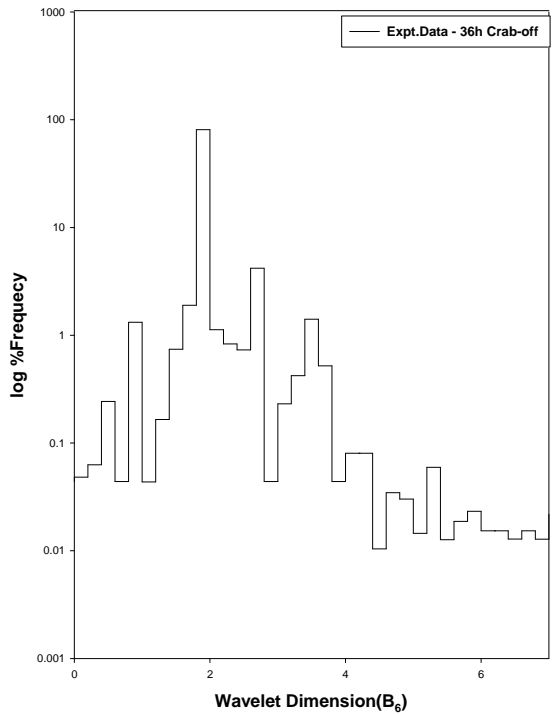}    
      \caption{\label{Fig.5}Wavelet dimension $B_{6}$ distribution of Cerenkov images recorded  by TACTIC telescope during $\sim$ 36h off-source scan.}
         \label{FigVibStab}
   \end{figure}

\begin{figure}
   \centering
     \includegraphics*[width=0.60\textwidth,angle=0,clip]{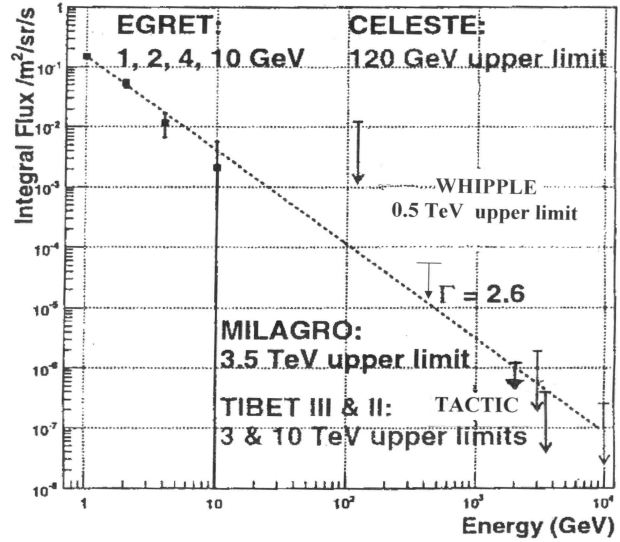}    
      \caption{\label{Fig.6}TACTIC $\geq$ 2 TeV  upper limits on the diffuse gamma-ray emission within galactic range of $|b|$$\leq$ -5$^{0}$ and $|l|$$\leq$ 200$^{0}$}
%         \label{FigVibStab}
   \end{figure}
%______________________________________________________________
\section{Conclusions}
It has been shown on the basis of a fractal and wavelet analysis of simulated atmospheric Cerenkov images, recorded in the TACTIC telescope, that it is possible to segregate cosmic ray proton-initiated events from diffuse gamma-ray initiated events with high efficiency using the wavelet dimension $B_{6}$  as the differentiating parameter. Application of data cuts based on this parameter have enabled us to provide a preliminary upper limit estimate of the diffuse galactic gamma-ray flux  above 2 TeV energy from a 36h off-source data set with the TACTIC telescope. We hope to improve the simulations  based results by optimizing the parameter cuts after more extensive simulations taking into account the cosmic ray spectrum and TACTIC characteristics.    
\section{Acknowledgements} 
%\label{}
The author is grateful to Ramesh Koul, Head Astrophysical Sciences Division, BARC for his kind support during the course of this work. Author is thankful to Dr.A.K.Razdan for his help in developing the software used in this work and to K.K.Yadav for the help rendered during the course of data analysis.


\begin{thebibliography}{}

\bibitem{}
G.F.Bignami,C.E.Fichtel, ApJ,(1974) L65
\bibitem{}
C.E.Fichtel, et al., ApJ, 198,(1975) 163.
\bibitem{}
S.Hayakawa, Phil  Mag. 43,(1952) 847.
\bibitem{}
S.D.Hunter et al., ApJ, 481,(1997) 205.
\bibitem{}
P.Sreekumar, BASI, 30, (2002) 61.
\bibitem{}
A.M.Hillas,Proc. 19$^{th}$  ICRC , La Jolla, 2 ,(1985) 445.
\bibitem{}
Aharonian,F.A. et al,Astron.$\&$ Astrophys.375,(2001),1008.
\bibitem{}
M.Amenomori et al.,Proc.25$^{th}$  ICRC, Durban , 3,(1997) 117.
\bibitem{}
S.LeBohec et al., ApJ, 539,(2000) 209.
\bibitem{}
Gus.Sinnis,Proc. 29$^{th}$ICRC, Pune ,3,(1997)101.
\bibitem{}
R.Atkins, et al,Physical Review Letters, 95,(2005) 251103 .
\bibitem{}
R.Koul, et al ,Nucl.Inst.Meth., A 578,(2007) 548.
\bibitem{}
C.L.Bhat, et al ,Nucl.Inst. Meth., A 340,(1994) 413.
\bibitem{}
J.N.Capdeville,et al,The Karlsruhe EAS Simulation Code 
CORSIKA,(1999)Report No.Kfk4998.
\bibitem{}
D.Heck, et al ,FZKA-report 6019, (1998) Forschunszentrum Karlsruuhe
\bibitem{}
D.Heck, et al , Astropart.Phys, 12 ,(1999) 145.
\bibitem{} 
Ong,Rene.A. 1998, Phys.Report , 305 , 93 .
\bibitem{}
M.K.Koul, et al , BASI(India), 30 ,(2002) 361.
\bibitem{}
A.K.Tickoo, Ph.D Thesis,University of Mumbai,(2002) India 
\bibitem{}
K.K.Yadav, et al ,Astropart.Phys,27,(2007) 447.
\bibitem{}
A.Haungs, et al, Astropart.phys., 12,(1999) 145.
\bibitem{}
A.Haungs,\& J.Knapp ,Proc. 27$^{th}$ ICRC ,Hamburg,1,(2001)315.
\bibitem{}
A.Haungs,,J.Knapp,I.Bond \& R.Pallassini, Proc.27$^{th}$ ICRC,Hamburg,
7,(2001)2910.
\bibitem{}
M.Amenomori et al, ApJ, 580 ,(2002) 887.
\bibitem{}
C.K.Bhat, et al, PS3,16$^{th}$ NSSS(India),Rajkot,(2010),74

\end{thebibliography}
\end{document}